\documentclass[10pt, conference, compsocconf]{IEEEtran}

\usepackage{xspace}
\usepackage[T1]{fontenc}
\usepackage{color}
\usepackage[hyphens]{url}
\usepackage{amsmath}
\usepackage{amssymb}

\newcommand{\PFLOPS}{1.54}

\newcommand{\partCPUtm}{Intel\textsuperscript{\textregistered} Xeon Phi\textsuperscript{\texttrademark}\xspace}
\newcommand{\partCPU}{Intel Xeon Phi\xspace}

\usepackage{graphicx}
\graphicspath{{figures/}}
\DeclareGraphicsExtensions{.pdf,.jpeg,.png}

\usepackage{pgfplots}
\pgfplotsset{compat=1.13}
\usepgfplotslibrary{colorbrewer}
\pgfplotsset{cycle list/Set1-4}

\usepackage{etoolbox}
\apptocmd{\sloppy}{\hbadness 10000\relax}{}{}

\begin{document}

\title{Cataloging the Visible Universe through Bayesian Inference at Petascale}

\author{Jeffrey~Regier\IEEEauthorrefmark{1},
Kiran~Pamnany\IEEEauthorrefmark{2},
Keno~Fischer\IEEEauthorrefmark{3},
Andreas~Noack\IEEEauthorrefmark{4},
Maximilian~Lam\IEEEauthorrefmark{1},
Jarrett~Revels\IEEEauthorrefmark{4},\\
Steve~Howard\IEEEauthorrefmark{5},
Ryan~Giordano\IEEEauthorrefmark{5},
David~Schlegel\IEEEauthorrefmark{6},
Jon~McAuliffe\IEEEauthorrefmark{5},
Rollin~Thomas\IEEEauthorrefmark{6},
Prabhat\IEEEauthorrefmark{6}\\\\
\IEEEauthorblockA{\IEEEauthorrefmark{1}Department of Electrical Engineering and Computer Sciences, University of California, Berkeley}
\IEEEauthorblockA{\IEEEauthorrefmark{2}Parallel Computing Lab, Intel Corporation}
\IEEEauthorblockA{\IEEEauthorrefmark{3}Julia Computing}
\IEEEauthorblockA{\IEEEauthorrefmark{4}Computer Science and AI Laboratories, Massachusetts Institute of Technology}
\IEEEauthorblockA{\IEEEauthorrefmark{5}Department of Statistics, University of California, Berkeley}
\IEEEauthorblockA{\IEEEauthorrefmark{6}Lawrence Berkeley National Laboratory}
}

\maketitle

\begin{abstract}
Astronomical catalogs derived from wide-field imaging surveys are an important tool for understanding the Universe.  We construct an astronomical catalog from 55~TB of imaging data using Celeste, a Bayesian variational inference code written entirely in the high-productivity programming language Julia.  Using over 1.3 million threads on 650,000 \partCPU cores of the Cori Phase~II supercomputer, Celeste achieves a peak rate of \PFLOPS\ DP PFLOP/s. Celeste is able to jointly optimize parameters for 188M stars and galaxies, loading and processing 178~TB across 8192 nodes in 14.6 minutes. To achieve this, Celeste exploits parallelism at multiple levels (cluster, node, and thread) and accelerates I/O through Cori's Burst Buffer. Julia's native performance enables Celeste to employ high-level constructs without resorting to hand-written or generated low-level code (C/C\texttt{++}/Fortran), and yet achieve petascale performance.
\end{abstract}

\begin{IEEEkeywords}
astronomy, Bayesian, variational inference, Julia,
high performance computing
\end{IEEEkeywords}

\section{Introduction}
\label{intro}
Astronomical surveys are the primary source of information about the Universe
beyond our solar system.  They are essential for addressing key open questions
in astronomy and cosmology about topics such as the life-cycles of stars and
galaxies, the nature of dark energy, and the origin and evolution of the
Universe.

The principal products of astronomical imaging surveys are catalogs of light sources,
such as stars and galaxies. These catalogs are generated by identifying
light sources in survey images and characterizing each according to physical
parameters such as brightness, color, and morphology. Astronomical catalogs are
the starting point for many scientific analyses, such as theoretical modeling of
individual light sources, modeling groups of similar light sources, or modeling
the spatial distribution of galaxies.  Catalogs also inform the design and
operation of follow-on surveys using more advanced or specialized
instrumentation (e.g., spectrographs). For many downstream analyses, accurately
quantifying the uncertainty of parameters' point estimates is as important as
the accuracy of the point estimates themselves.

\begin{figure}[t]
\centering
\includegraphics[width=3in]{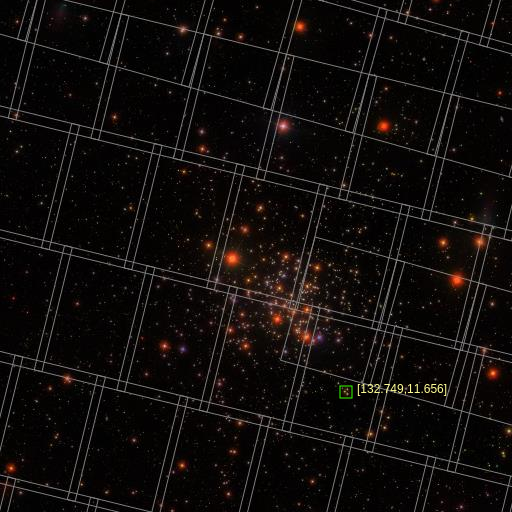}
\caption{SDSS data highlighting image boundaries. Some images overlap
substantially. Some light sources appear in multiple images that do not overlap.
Celeste uses all relevant data to locate and characterize each light source
whereas heuristics typically ignore all but one image in regions with overlap.
Credit:~SDSS DR10 Sky Navigate Tool.\vspace{-10px}}
\label{sdss_fields}
\end{figure}

Modern astronomical surveys produce vast amounts of data. Our work uses
images from the Sloan Digital Sky Survey (SDSS)~\cite{dr12} as a test case to
demonstrate a new, highly scalable algorithm for constructing astronomical
catalogs. SDSS produced 55~TB of images covering 35\% of the sky and a catalog
of 470~million unique light sources. Figure~\ref{sdss_fields} shows image
boundaries for a small region from SDSS. Advances in detector fabrication
technology, computing power, and networking have enabled the development of a
new generation of upcoming wide-field sky surveys orders of magnitude larger than SDSS. In 2020, the Large Synoptic Survey Telescope (LSST) will
begin to obtain more than 15~TB of new images nightly~\cite{LSST}. The
instrument will produce 10s--100s of PBs of imaging and catalog data over the
lifetime of the project.

Producing an astronomical catalog is challenging in part because the
parameters of overlapping light sources must be learned collectively: the
optimal parameters for one light source depend on the optimal parameters of
nearby light sources, and vice versa.

This paper presents Celeste, a novel computer program for inferring astronomical
catalogs. Celeste makes three major advances related to high-performance data
analytics:
\begin{enumerate}
\item Celeste is the largest reported application to date of
variational inference, an approximate Bayesian inference technique, to a dataset
from any domain.
\item Celeste demonstrates that Julia~\cite{bezanson2017julia}, a
high-productivity programming language, excels at HPC, achieving
performance previously attained only by languages like C, C\texttt{++} and
Fortran.
\item Celeste is the first program to perform a fully Bayesian analysis of an
entire modern astronomical imaging survey. It significantly improves on the
accuracy and uncertainty quantification of its predecessors.
\end{enumerate}

Section~\ref{background} provides additional context for each of the three
advances above.
Section~\ref{model} describes the statistical model that we adopt for this work, and derives a numerical optimization problem that amounts to applying
this model to data.
Section~\ref{distributed} describes a distributed numerical optimization strategy that lends itself to effective scaling.
Section~\ref{julia} evaluates the suitability of the Julia programming language for HPC.
Section~\ref{measurement} discusses our performance measurement methodology.
Section~\ref{results} reports on weak scaling, strong scaling, and a performance
run that attained \PFLOPS{} PFLOP/s.
Section~\ref{science} assesses the scientific quality of the catalog. Large
improvements on several metrics are reported over the current standard practice.
Section~\ref{discussion} considers the broader implications of our results
for three research areas: Bayesian inference, programming languages, and
astronomy.

\section{Background}
\label{background}
Constructing an astronomical catalog is computationally demanding for any
algorithm because of the sheer size of the datasets. Thus, approaches to date
have been largely based on computationally efficient heuristics rather than
Bayesian methods~\cite{lupton2005sdss, bertin1996sextractor}. Heuristics have a
number of drawbacks: 1)~they do not make optimal use of prior information
because it is unclear how to ``weight'' it in relation to new information; 2)~
they do not effectively combine knowledge from multiple image surveys, or even
from multiple overlapping images from the same survey; and 3)~they do not
correctly quantify uncertainty of their estimates. They may flag some estimates
as particularly unreliable, but confidence intervals follow only from a
statistical model.  Without modeling the arrival of photons as a Poisson
process, for example, there is little basis for reasoning about uncertainty in
the underlying brightness of light sources.

These shortcomings are addressed by the Bayesian formalism.  An astronomical
catalog's entries are modeled as unobserved random variables with prior
distributions.  The posterior distribution encapsulates knowledge about the
catalog's entries, combining prior knowledge of astrophysics with new data from
surveys in a statistically efficient manner; it represents the uncertainty.
Because most light sources will be near the detection limit, these uncertainty
estimates are as important as the parameter estimates themselves for many
analyses.

Unfortunately, exact Bayesian posterior inference is NP-hard for most
probabilistic models of interest~\cite{bishop2006pattern}. Approximate Bayesian
inference is an area of active research. Markov chain Monte Carlo (MCMC) is the
most common approach. Unfortunately, the computational work required to draw
enough ``samples'' makes it poorly suited to large-scale problems. It is also
difficult to determine when the Markov chain has mixed.

To date, Tractor~\cite{thetractor} is the only program for Bayesian posterior
inference applied to a complete modern astronomical imaging survey. Tractor is a
single-threaded program written in Python. It relies on Laplace approximation, in which
the posterior is approximated with a multivariate
Gaussian distribution centered at the mode, with the Hessian of the log
likelihood function at the mode as its covariance matrix. This type of
approximation is not suitable for categorical random variables, as no Gaussian
distribution could approximate them well. Additionally, because
Laplace approximation centers the Gaussian approximation at the mode rather than
the mean, the solution depends heavily on the parameterization of the
problem~\cite{bishop2006pattern}. Nonetheless, Laplace approximation has been
successfully applied to large-scale
problems~\cite{alexanderian2016fast}.

Variational inference (VI) extends Laplace approximation to approximate posterior
distributions with non-Gaussian distributions. Like Laplace approximation, it
uses numerical optimization to find a distribution that approximates the
posterior without sampling~\cite{blei2017variational}. In practice, the
resulting optimization problem is often orders of magnitude faster to solve
compared to MCMC approaches. Scaling VI to large datasets is nonetheless challenging.
The largest published
applications of VI to date have been to text mining, where topic models are fit to
several gigabytes of text~\cite{hoffman2013stochastic}. Modern astronomical
datasets are at least four orders of magnitude larger than that.

\subsection*{Julia for HPC}
To date, most HPC programs have been written in relatively verbose, low-level
languages like C, C\texttt{++}, or Fortran. On the other hand, high-level
languages' expressiveness, feature-richness, and succinctness have motivated
practitioners to develop techniques which mitigate their computational overhead.
These techniques usually subscribe to Ousterhout's two-language paradigm: one
should use a systems programming language to write optimized kernels for
performance ``hot spots'', and use a scripting language to ``glue'' the kernels
together and steer the overall computation~\cite{ousterhout1998scripting}.
NumPy~\cite{walt2011numpy} and PyFR~\cite{vincent2016pyfr} are examples of Ousterhout's paradigm.

While the two-language approach has made leveraging high-level languages for
HPC more feasible, it features a notable drawback: a sizable fraction of
scientific computing projects begin with domain experts prototyping code in a
high-level language, with the assumption that further in the development
timeline, HPC experts will be able to identify performance bottlenecks and
rewrite them into fast kernels by porting the code to a low-level language.
Porting is time-consuming, expensive, error-prone, and can
prevent rapid iteration between HPC experts and domain experts.

Julia is a high-level programming language that nonetheless lets programmers
attain high performance. By leveraging just-in-time (JIT) compilation, multiple
dispatch, type inference, and LLVM~\cite{lattner2004llvm}, Julia enables
rapid, interactive prototyping while achieving performance competitive with C,
C\texttt{++}, and Fortran~\cite{bezanson2017julia}. Julia accomplishes this
without forcing the programmer to use third-party  ``accelerators'' (e.g. Numba,
PyPy) or requiring hot kernels to be written in a low-level language.

Julia code is typically 5x to 10x shorter than code implementing the same
functionality in C, C++, or Fortran. Computationally efficient Julia code is
only slightly less succinct. Using a computationally efficient
``StaticArray'', for example, is only slightly more verbose than a native
(mutable) array. Moreover, even in an HPC codebase, the vast majority of
the code accounts for little of the runtime. Here Julia is at least as succinct
as Python.

\section{The Statistical Model}
\label{model}
\begin{figure}[!t]
\centering
\includegraphics[width=2in]{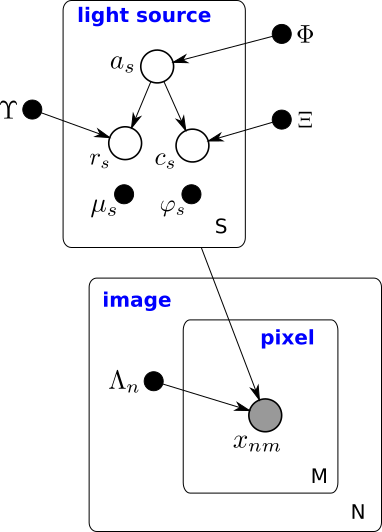}
\caption{A graphical representation of the model we implement for astronomical
images. Shaded vertices represent observed random
variables. Empty vertices represent latent random variables.  Black dots
represent constants. Constants denoted by uppercase Greek characters are fixed
throughout our procedure. Constants denoted by lowercase Greek
letters are inferred, along with the posterior distribution of
the latent random variables. Edges signify permitted conditional
dependencies. The rounded rectangles represent independent replication.\vspace{-5px}}
\label{graphical_model}
\end{figure}

We adopt the probabilistic model from~\cite{regier2015celeste}, which reports
state-of-the-art scientific results but only scales inference to a several
hundred stars and galaxies. The model is a joint probability distribution over
observed random variables (the pixel intensities) and unobserved, or latent,
random variables (the catalog entries). It is represented graphically in
Figure~\ref{graphical_model}.

For image $n=1,\ldots,N$, $\Lambda_n$ is a constant vector of metadata
describing its sky location and the atmospheric conditions at the time of the
exposure.  Each pixel $m=1,\ldots,M$ in image $n$ has an intensity $x_{nm}$.
This intensity is an observed random variable following a Poisson distribution
with a rate parameter $F_{nm}$ unique to that pixel. $F_{nm}$ is a deterministic
function of the catalog (which includes random quantities) and the image
metadata.  Pixel intensities are conditionally independent given
the catalog.

For light source $s=1,\ldots,S$, the Bernoulli random variable $a_s$ indicates
whether it is a star or a galaxy; a log-normal random variable $r_s$ denotes its
brightness in a particular band of emissions (the ``reference band''); and a
multivariate normal random vector represents its colors---defined formally as
the log ratio of brightnesses in adjacent bands, but corresponding to the
colloquial definition of color. These random quantities have prior distributions
with parameters $\Phi$, $\Upsilon$, and $\Xi$, respectively. These parameters
are learned from preexisting astronomical catalogs.  Additionally, each light
source is characterized by unknown constant vectors $\mu_s$ and $\varphi_s$. The
former indicates the light source's location and the latter, if the light source
is a galaxy, represents its shape, scale, and morphology. Though non-random,
these vectors are nonetheless learned within our inference procedure.

Our primary use for the model is
computing the distribution of its unobserved random variables conditional on a
particular collection of astronomical images. This distribution is known
as the posterior, and denoted $p(z | x)$, where
$x := \{x_{nm}\}_{n=1,m=1}^{N,M}$ represents all the pixels and
$z := \{a_s, r_s, c_s\}_{s=1}^S$ represents all the latent random variables.
Exact posterior inference is computationally intractable for
the proposed model, as it is for most non-trivial Bayesian models.

Variational inference (VI) finds an approximation to the posterior
distribution $p(z | x)$ from a class of candidate distributions through numerical
optimization~\cite{blei2017variational}.
The candidate approximating
distributions $q_\theta$ are parameterized by a real-valued vector $\theta$. VI
maximizes (with respect to $\theta$) the following lower bound on the log probability of the data:
\begin{align}
\log p(x_{11},\ldots,x_{NM})
&\ge \mathbb{E}_{q_\theta} \left[ \log p(x, z) - \log q_\theta(z) \right] \label{elbo}\\
&=: \mathcal L(\theta).
\end{align}
This lower bound holds for any $q_\theta$, by Jensen's inequality.
We restrict $q_\theta$ to a class of distributions that let us evaluate
the expectation in Equation~\ref{elbo} analytically. A complete description
appears in~\cite{regier2015celeste}.

To date, this approach to inference has only been applied to a small astronomical
dataset containing 654 stars and galaxies~\cite{regier2015celeste}. In that
work, numerical optimization is single-threaded, and light sources parameters
are optimized in isolation, rather than jointly over the parameters
for all the light sources. Joint optimization is a necessity for highly accurate scientific results.

\section{Extreme scaling}
\label{distributed}

Inferring an astronomical catalog for the SDSS dataset with the proposed
statistical model amounts to solving a massive constrained optimization problem.
Each of the 469 million celestial bodies has 44 parameters. Stored in double
precision, the parameters alone are 154 GB. Moreover, each celestial body cannot
be optimized in isolation: the optimal setting of its parameters depends on the
optimal setting of nearby celestial bodies' parameters, and vice versa. Yet the
vast majority of pairs of celestial bodies can be optimized independently of
each other (Figure~\ref{sdss_fields}).

Approaches to distributed optimization like HogWild~\cite{recht2011hogwild} show
that complicated locking mechanisms are not necessary to attain convergence to a
stationary point.  But those results apply only to stochastic gradient descent,
which converges in sublinear time in theory and requires tens of thousands of
iterations in practice. Newton's method, on the other hand, converges
quadratically in theory and in tens of iterations in practice, by leveraging
second-order information. Because each iteration necessitates visiting a
substantial fraction of the 55 TB dataset, exploiting second-order information
is essential.

\begin{figure}[t]
\includegraphics[width=\columnwidth]{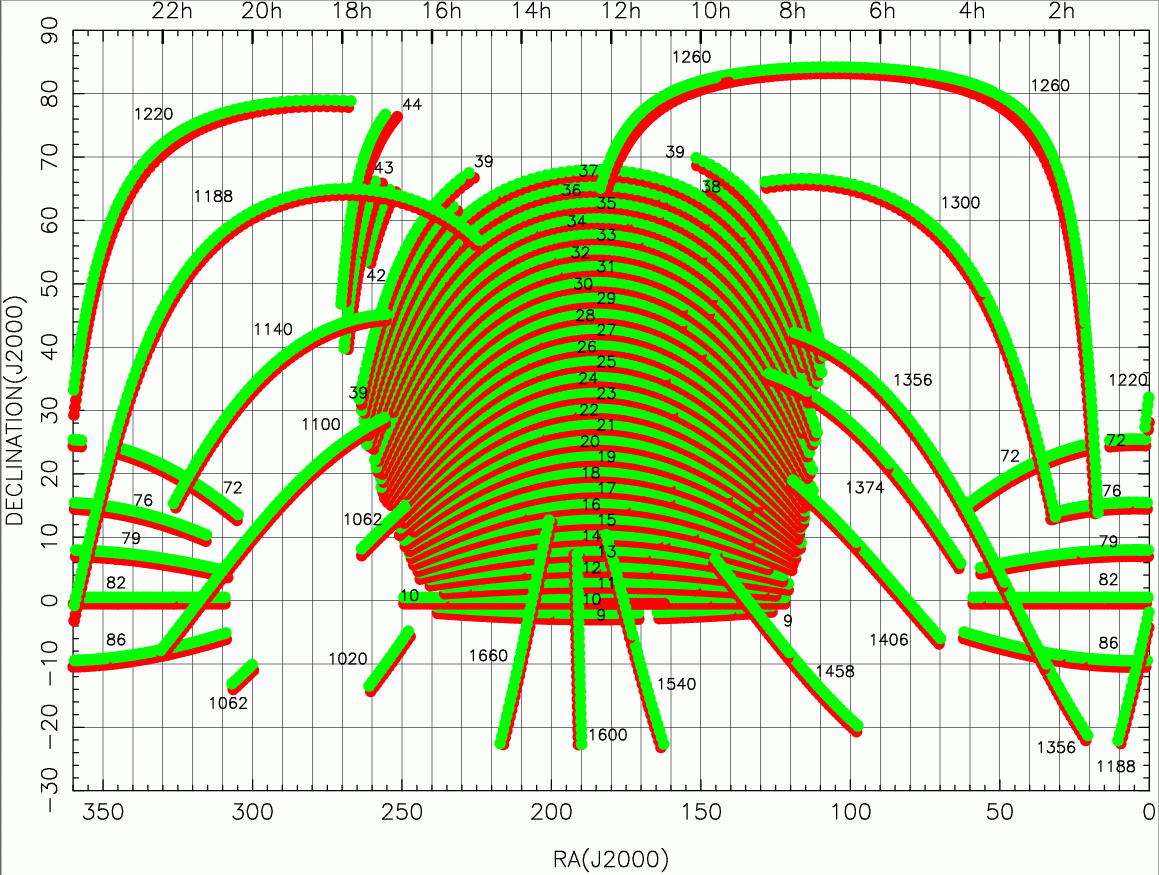}
\caption{SDSS sky coverage map. Each monocolor arc (a ``stripe'') represents the sky
photographed during a particular night. The telescope remains fixed while the
Earth rotates. Axes units are degrees of right ascension (RA) and declination
(DEC) with respect to the equator. Credit: SDSS.\vspace{-10px}}
\label{sdss_coverage}
\end{figure}

Our method has three levels: 1) non-overlapping rectangular regions of the sky
are optimized concurrently on different nodes; 2) light sources within the same
region of sky are optimized in parallel by different threads if they do not
overlap and 3) individual light sources are optimized by a vectorized
implementation of a variant of Newton's method for nonconvex optimization.
The outer- and mid-level of the optimization procedure correspond to block
coordinate ascent, while the lowest level is second-order optimization. At each
level, the parameters for light sources not optimized are held constant.

The description above explains why our optimizer converges, and gives some
indication of how different routines align with the hardware. Now we
give a more algorithmic description.

\subsection{Task decomposition} \label{decomposition}
The SDSS imaging camera scans the sky in stripes along ``great circles''
(Figure~\ref{sdss_coverage}). Each stripe is divided into 12 MB image files that
are stored on disk (Figure~\ref{sdss_fields}). Stripes overlap, and different
stripes were scanned a differing number of times. To reason about a particular
light source, all the images of it need to be loaded. The number varies
significantly---between 5 and 480 images (up to 5.5 GB).

Most images contain hundreds or thousands of light sources. It is efficient to
process multiple nearby light sources together on the same compute node so the
images containing them just need to be loaded once for all of them, rather than
once for each of them. Additionally, nearby light sources need to be optimized
jointly. Optimizing them together on the same node allows for communication
among threads through shared memory rather than over the network.

Therefore, we define our node-level tasks to correspond to contiguous regions of
sky. Each task is to (jointly) find optimal parameters for the light sources
in a particular region of the sky, with light sources in neighboring regions
held fixed.

If our tasks correspond to large regions of sky, then most images that overlap
with the region will lie entirely within it. That is desirable: the images on
the boundary of a region typically overlap with another region too. We can
minimize the mean number of times an image is loaded by specifying
large tasks.

However, large tasks create inefficiencies both at the beginning and the end
of a job. At the beginning of a job, the first task for each compute node cannot
start processing until the image data is loaded. For subsequent tasks, the
nodes can prefetch images before the previous task has completed, but not for
the first task. At the end of a job, once the queue of unclaimed tasks is
empty, some compute nodes will remain idle while others are still finishing
their last task. We refer to this as load imbalance. If the tasks are long
running, the load imbalance will be worse.

This poses a trade-off between effective load balance and image loading
overhead. Smaller tasks allow for more effective load balance, but the same
images must be loaded repeatedly. Larger tasks reduce the I/O burden, but
simultaneously increase the load imbalance between processes.

It is not enough to partition the sky into uniformly sized regions, sized
optimally to balance the trade-off between redundant image loading and load
imbalance. No uniform size is good. Uniformly sized regions of the sky vary
significantly in the number of light sources they contain, and the number of
images of those light sources. Highly irregular task durations leads to high
load imbalance.

Instead, we leverage an existing astronomical catalog to generate our tasks. We
partition the sky recursively into regions that we expect to contain roughly the
same number of bright pixels, based on existing astronomical catalogs. Bright
pixels correlate with the amount of processing that will subsequently be needed.
Task generation does not require loading image data, just an existing catalog,
so we do it during preprocessing, as a one-off job, executed on a small number
of nodes.

A task description also lists the light sources in the region to optimize
subsequently, and gives initial values for these light sources' parameters,
derived from existing astronomical catalogs. Thus, it is an existing
astronomical catalog that determines $S$, the total number of light sources in
our statistical model. We may ultimately develop our own scripts for proposing
candidate light sources for future datasets; for SDSS, however, existing
catalogs work well for initializing our algorithm.\vspace{5px}

Tasks may be executed simultaneously: they consider non-overlapping regions of
sky, so they are nearly independent in our objective function. However, light
sources near a border of a region may not reach their optimal value if a
neighboring region contains light sources that contribute photons to both
region. For example, if a region contains a large, bright galaxy that extends
beyond that region's border, but that galaxy's current parameters are not
optimal, then the predictions in neighboring regions could also be off to
compensate.

We address this by introducing a second stage to our algorithm---creating a
second partitioning of the sky by shifting each region in the first partition by
a fixed amount. Light sources near a border in the first partition
will almost always away from a border in the second partition.\footnote{A third partition may be needed to completely optimize light sources near a ``corner''---at the intersection of boundaries from the first and second partitions. In practice, improvements are negligible from introducing more than two partitions.}

Each
such partition constitutes a stage; tasks for the second stage are processed
only after all tasks for the first stage are complete.

Task generation is the only stage of our algorithm completed during
preprocessing. All other steps are part of the main job that we benchmark.

\subsection{Task scheduling}
Although we generate tasks to have roughly the same computational difficulty,
their durations nonetheless vary too much for static scheduling. We use
Dtree~\cite{pamnany2015dtree}, a distributed dynamic scheduler that balances
load for irregular tasks, even at petascale. Dtree organizes compute nodes into
a tree whose height scales logarithmically in the number of nodes. To distribute
tasks, each node only needs to communicate with its parent and its immediate
children.

\subsection{Shared state}
During the optimization procedure, the current parameters for all celestial
bodies are stored in a partitioned global address space (PGAS)~\cite{pgas}. Our
interface mimics that of the Global Arrays Toolkit~\cite{nieplocha2006advances}. We
use MPI-3 as the transport layer; get and put operations on elements make use of
one-sided RMA operations that are supported in hardware on most supercomputer
fabrics~\cite{hoefler2015rma}.

\subsection{Task processing} \label{task-processing}
Each node runs several multi-threaded processes. Each process receives a task
for processing from the scheduler and performs some initialization
steps---determining the relevant images to load as well as loading them from
storage, fitting some image-specific parameters, and retrieving any previously
computed parameters for light sources in the region. A typical task involves
jointly optimizing roughly 500 light sources, each characterized by 44
parameters (about 20k parameters to optimize in total per task).

Multiple threads then coordinate to jointly optimize the light sources for
the current task. They employ block coordinate ascent; each light source's
44 parameters form a block. At each step of the algorithm, one light source's
parameters are optimized to machine tolerance by Newton's method, with step
sizes controlled by a trust region~\cite{wright1999numerical}. The trust
region ensures convergence to a stationary point from any starting point
even though the objective function is, in general, nonconvex.

Block coordinate ascent is a serial algorithm: if multiple overlapping blocks
(of parameters) are updated in parallel the algorithm no longer converges in
general. Instead, for a region of sky, threads coordinate their work through the
Cyclades approach to conflict-free asynchronous machine
learning~\cite{pan2016cyclades}. Cyclades bases thread assignments on a
``conflict graph.'' Nodes are light sources and edges indicate a conflict. Light
sources are in conflict if they overlap. Light sources that conflict cannot be
optimized concurrently. At each iteration, Cyclades samples light sources at
random without replacement and partitions the sample into connected components,
according to the conflict graph restricted to the sample. Then, connected
components are distributed among threads; lights sources that overlap in the
sample are all assigned to the same thread. Cyclades works well because even if
the conflict graph is connected, its restriction to a random sample of nodes
typically has many connected components.

Each thread optimizes a particular light source's parameters with any
overlapping light sources' parameters held fixed. By using Newton steps with
exact Hessians rather than L-BFGS or a first-order optimization method, we
attain a 1--2 order-of-magnitude speed-up. While L-BFGS is a robust and widely
used optimization method, it struggles with the objective function for our
problem, taking up to 2000 iterations to converge. Newton's method, on the other
hand, converges reliably on our problem in tens of iterations. Because the
optimization problem is nonconvex, we also introduce a trust region constraint,
as detailed in~\cite{wright1999numerical}. To apply Newton's method, exact
Hessians must be computed at each iteration. Each Hessian is a dense symmetric
matrix with 44 rows and 44 columns. Computing the Hessian along with the gradient,
rather than the gradient alone, takes 3x longer, but can reduce the total number
of iterations by 100x relative to L-BFGS (which only requires gradients).

\section{PetaFLOP performance with Julia}
\label{julia}
Our implementation uses Julia's multi-threading capabilities and Lisp-like
macros, as well as external pure-Julia implementations of statically sized
arrays~\cite{staticarrays}. Additionally, we use a Julia package for performing
array-of-structs to struct-of-arrays conversion~\cite{structsofarrays}, enabling
the compiler to replace certain expensive gather instructions with simple loads.

We also use automatic differentiation (AD), both
forward-mode~\cite{revels2016forwarddiff} and reverse-mode~\cite{reversediff},
but only where exploiting the sparsity of the Hessian  is not required for good
performance. While advanced sparsity-exploiting AD tools have been  developed in
Julia~\cite{feng2016sparsehessian}, these tools are
either too experimental or restrictive for our use case. Instead, for
performance-critical  code, we use our own hand-coded derivatives that leverage
custom index types to exploit  Hessian sparsity structure.

During our final push toward attaining petascale performance, we made two types
of changes: source-level refactoring of Celeste, and enhancements to the
compiler in order to improve the quality of the generated instructions. For the
former, Celeste's code was refactored to reduce dynamic memory allocations,
avoid dynamic method lookup, exploit sparsity patterns in matrix operations,
ensure iteration order matched memory layout order, and reduce the memory
footprint of per-pixel computations to fit almost entirely into L1 cache.

Additionally, we made three key improvements to Julia's compiler. First,
Julia's ability to recognize and evaluate at compile time certain constant
indexing computations was improved by increasing the precision of various
analyses within type inference. Second, dereferenceability and automatic
aliasing information was added to generated LLVM IR, augmented by user-provided
code annotations and loop unrolling hints exposed as Julia macros. Third, LLVM
compilation passes were improved by performing cost model adjustments,
experimenting with pass order, and fixing instances where aliasing information
was incorrectly discarded during compilation. Once these changes are officially
incorporated upstream, they can benefit all Julia users and users of other LLVM
frontends, such as the Clang C\texttt{++} compiler.

With these improvements, the compiler's vectorization and code generation
capabilities---in particular, the fusion of multiplications and additions---allowed
Celeste to obtain excellent code quality without resorting to low-level
languages, assembly intrinsics, or handwritten libraries. Since most of these
improvements were not target architecture specific, they yielded improvements
not only on Celeste's primary target architecture (\partCPU), but
also on other architectures (notably, Haswell).

\section{Performance Measurement}
\label{measurement}

\subsection{System} \label{system}
We measure Celeste's performance on the Cray~XC40 Cori supercomputer (``Phase~II'') at the National Energy Research Scientific Computing Center (NERSC).
Cori Phase~II contains 9688 compute nodes, each with one 1.40GHz 68-core \partCPUtm 7250 processor.\footnote{Intel, Xeon, and \partCPU are trademarks of Intel
Corporation in the U.S. and/or other countries.}
Each core includes two AVX-512 vector processing units.
Cores on a node connect through a two-dimensional mesh network with two cores per ``tile.''
Within a tile, cores share a 1~MB cache-coherent L2 cache.

Each node has 96~GB of DDR4 2400~MHz memory using six 16~GB DIMMs (115.2 GB/s peak bandwidth).
Each node also includes 16~GB of on-package high-bandwidth memory with bandwidth five times that of DDR4 DRAM ($>460$ GB/sec).
Cori's high speed network is a Cray Aries high speed ``dragonfly'' topology interconnect.

Campaign data are stored on Cori's 30~PB Lustre file system, with a theoretical aggregate bandwidth exceeding 700~GB/s.
In addition, Cori features a Burst Buffer, a layer of non-volatile storage that sits between the processors' memory and the parallel file system that serves to accelerate I/O performance of applications \cite{Bhimji2016AcceleratingSW}.
Cori includes 280 Burst Buffer nodes.
Each of these contains two Intel P3608 3.2~TB NAND flash SSD modules attached over two PCIe gen3 interfaces.
The Cori Burst Buffer provides approximately 1.7~TB/s of peak I/O performance with 28M~IOPs, with 1.8~PB storage capacity.




\subsection{Measurement methodology}
\label{flops}

All floating point operations (FLOPS) performed by Celeste are in double
precision.
The main source of FLOPS is evaluating the objective function
and its gradient and Hessian. These evaluations' FLOP counts are proportional to the
number of active pixels. We use the Intel Software Development Emulator
(SDE) to measure one and two active pixel visits and determined that each
active pixel visit performs 32,317 FLOPS. We then count the number of
active pixel visits during a run in order to measure total FLOPS.

The evaluation of the objective functions is the primary but not the
only source of FLOPS. Other sources include the Newton trust region
algorithm, as our implementation computes an eigen decomposition, as well
as several Cholesky factorizations at each iteration. We again use SDE,
this time for a larger multi-threaded run on a single node, and compare
the flop count measured with the flop count derived from active pixel
visits. We find that these additional sources of FLOPS increase the total
flop count to 1.375 times the FLOP count derived from active pixel
visits alone.

Thus, we determine the total number of FLOPS for Celeste by counting
the number of active pixel visits during a run, and applying {the 1.375 factor}
to account for FLOPS outside the objective function.

\section{Performance Results}
\label{results}

\subsection{Per-thread analysis}\label{per-thread}
Each thread's runtime is 67\% Julia generated code, 18\% native dependencies
such as image I/O libraries and the Julia runtime (memory allocation, thread
management, etc.), 10\% the system math library, 3\% the Intel Math Kernel
Library, and 2\% the C standard library and the Linux kernel. Of all FLOPs,
82.3\% operated on 8-wide AVX512 vector registers.

\subsection{Per-node analysis}
We empirically determined that eight threads per process and 17 processes per
\partCPU processor yields the highest throughput. Changing the number of threads
and processes modifies memory access patterns in complicated ways as several factors are in tension: with more threads,
several may remain idle at the end of a task while the
last few light sources are optimized. With fewer threads, tasks take longer to
complete, leading to more processes remaining idle while other processes finish
their last task. An additional consideration is that threads share memory,
whereas processes do not. Empirically, the 8x17 (threads x processes) node configuration does well at balancing all these considerations.

\subsection{Multi-node scaling}
We assess both weak and strong scaling with as many as 8192 nodes (557,056 CPU
cores). The list of tasks is generated during preprocessing, as explained in
Section~\ref{decomposition}. Every other step of our algorithm is measured as
part of these scaling runs. We breakout runtimes into four components.
\begin{enumerate}
\item \textit{image loading} -- The time taken to load images while worker threads
are idle. This time is accumulated by each process only for the first task.
Images for subsequent tasks are prefetched.
\item \textit{load imbalance} -- All processes except the last one to finish accumulate
load imbalance time after completing their last task; they are idle because
the dynamic scheduler has no unassigned tasks left.
\item \textit{task processing} -- The main work loop, detailed in
Section~\ref{task-processing} and profiled in section~\ref{per-thread}.
This component involves no network or disk I/O: only computation and shared memory access.
No time when threads are idle is included in this component.
\item \textit{other} -- Everything else, always a small fraction of the total runtime.
This include scheduling overhead, all network I/O (excluding image loading), and
writing output to disk.
\end{enumerate}

For weak scaling, the problem size scales with the number of nodes, whereas for
strong scaling, the problem size remains fixed. But how do we measure the size of a particular problem? If we double the square
degrees of sky to process, we may end up increasing the problem difficulty by
more or less than 2x: some regions are much more dense with bright light sources
than others, or are imaged more times, as touched on in
Section~\ref{decomposition}. It is tempting to consider just regions of sky that have
roughly the same density of images, light sources, and bright light sources.
But then our problem is not a single contiguous region of sky. Discontiguous
regions have fewer dependencies between tasks, and these dependencies are of
primary interest for multi-node scaling experiments.

Our solution is to treat the number of tasks, as determined by the preprocessing
procedure in Section~\ref{decomposition}, as the size of the problem, and to
use only contiguous regions of sky as problems. Tasks are recursively subdivided
regions of sky that have roughly equal work. While runtimes for them can vary,
that variation is an essential aspect of our approach, and
thus something to preserve in our scaling runs.

For the largest scaling runs (139,264 processes), the size of our test dataset
(SDSS) limits us to just four tasks per process. That gives the ``image
loading'' component and the ``load imbalance'' components exaggerated
importance. That does not limit the usefulness of the scaling results, but it
needs to be accounted for when interpreting the scaling graphs.

\subsubsection{Weak scaling}
\begin{figure}[ht]
\begin{tikzpicture}
\begin{semilogxaxis}[
const plot,
smooth,
width=3.2in,
height=2.5in,
stack plots=y,
area style,
xlabel=Nodes (log scale),
xtick={2,8,32,128,512,2048,8192},
xticklabels={2,8,32,128,512,2048,8192}, 
enlarge x limits=false,
log basis x={2},
log ticks with fixed point,
ymin=0,
ymax=1300,
ylabel=Seconds,
legend pos=north west,
legend columns=1,
legend cell align={left},
legend style={font=\scriptsize}]

\addplot[fill=orange] table[x=nodes,y=opt_srcs] {figures/weak_scaling_GB.txt}
\closedcycle;
\addlegendentry{task processing}

\addplot[fill=cyan] table[x=nodes,y=load_wait] {figures/weak_scaling_GB.txt}
\closedcycle;
\addlegendentry{image loading}

\addplot[fill=magenta] table[x=nodes,y=wait_done] {figures/weak_scaling_GB.txt}
\closedcycle;
\addlegendentry{load imbalance}

\addplot[fill=gray] table[x=nodes,y=other] {figures/weak_scaling_GB.txt}
\closedcycle;
\addlegendentry{other}
\end{semilogxaxis}
\end{tikzpicture}

\caption{Weak scaling. Runtime increases by 1.9X
between 1 node (68 cores) and 8192 nodes (557,056 cores)
while the number of tasks per node remains constant.
See accompanying discussion of load imbalance.
\vspace{-5px}}
\label{fig:weak_scaling}
\end{figure}
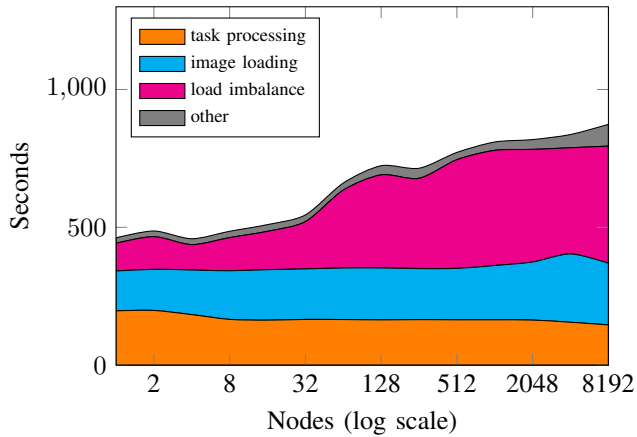

To assess weak scaling, we complete 68 tasks per node, or, equivalently, 4 tasks
per process.
Figure~\ref{fig:weak_scaling} shows runtimes for 1 node through 8192 nodes.

Task processing runtime is nearly constant throughout this range. This is expected since task processing is performed without communication.

Image loading time is also constant as the number of nodes grows---in part a testament
to the hardware. Images are staged on an SSD array and accessed over the high speed
interconnect described in Section~\ref{system}.

Load imbalance comes to dominate runtime past 32 nodes (4352 threads), but this
is more an artifact of the limited numbers of tasks (discussed above) than an
issue with the approach. With just four tasks per process, its no scheduler could keep all the processes busy during a substantial fraction of the job.
Nonetheless, the 8192 node
run completes all 557,056 tasks in about 15 minutes, finding optimal parameters
for 188,107,671 light sources.

\subsubsection{Strong scaling}
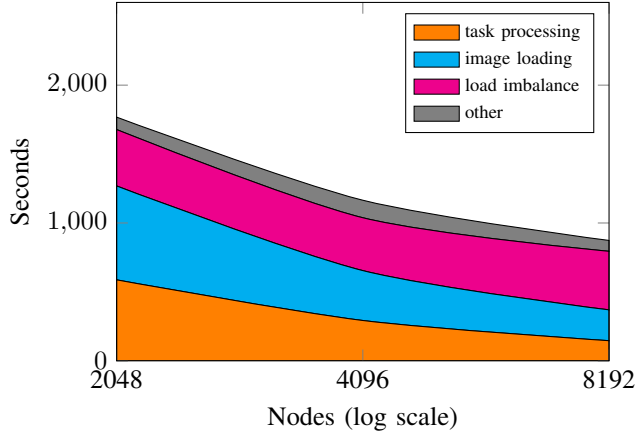
\begin{figure}[ht]
\begin{tikzpicture}
\begin{semilogxaxis}[
const plot,
smooth,
width=3.2in,
height=2.5in,
stack plots=y,
area style,
xtick={2048,4096,8192},
xticklabels={2048,4096,8192},
xlabel=Nodes (log scale),
enlarge x limits=false,
log basis x={2},
log ticks with fixed point,
ymin=0,
ymax=2600,
ylabel=Seconds,
legend pos=north east,
legend columns=1,
legend cell align={left},
legend style={font=\scriptsize}]

\addplot[fill=orange] table[x=nodes,y=opt_srcs]{figures/strong_scaling_GB.txt}
\closedcycle;
\addlegendentry{task processing}

\addplot[fill=cyan] table[x=nodes,y=load_wait]{figures/strong_scaling_GB.txt}
\closedcycle;
\addlegendentry{image loading}

\addplot[fill=magenta] table[x=nodes,y=wait_done]{figures/strong_scaling_GB.txt}
\closedcycle;
\addlegendentry{load imbalance}

\addplot[fill=gray] table[x=nodes,y=other]{figures/strong_scaling_GB.txt}
\closedcycle;
\addlegendentry{other}

\end{semilogxaxis}
\end{tikzpicture}

\caption{\textcolor{black}{Strong scaling. See accompanying discussion of load imbalance.}\vspace{-5px}}
\label{fig:strong_scaling}
\end{figure}

To evaluate strong scaling, we completed all 557,056 tasks with varying
numbers of nodes. Figure~\ref{fig:strong_scaling} shows runtimes and
their components at each scale.
Image loading and task processing
both exhibit near perfect scaling. ``Other'' remains constant as the node
count grows, but remains a small fraction of overall runtime.
Load imbalance grows in relative importance as the number of nodes increases.
Particularly at high node counts, the degree of load imbalance reflects how
few tasks are available per process.
For all components, we observe 65\% scaling efficiency from 2k to 4k
nodes, and 50\% efficiency from 2k to 8k nodes.

\subsection{Performance run}

Processes are not synchronized and may be in different states at various points
during a run---loading images, optimizing sources in a region of sky, or storing
results. In order to determine the true peak performance that can be
accomplished for Bayesian inference at scale, we prepared a specialized
configuration for performance measurement in which the processes synchronize
after loading images, prior to the optimization step. We then measure FLOPS
performed across all processes by counting active pixel visits
(Section~\ref{flops}) at one-minute intervals through the optimization step.

We ran this configuration on 9568 Cori \partCPU nodes, each running 17 processes
of eight threads each, for a total of 1,303,832 threads. 57.8 TB of image data was processed over a ten minute interval; the peak performance achieved was \PFLOPS\ PFLOP/s.

\begin{table}
\centering
\caption{Sustained flop rate}
\begin{tabular}{lrrr}
\hline
         & task processing & +load imbalance & +image loading \\ \hline
 TFLOP/s & 693.69  & 413.19          & 211.94 \\ \hline
\end{tabular}
\label{tbl:sustained}
\end{table}

We report on sustained performance in Table~\ref{tbl:sustained}, using the
standard Celeste configuration on 9600 nodes. This run completed 326,400
tasks in about seven minutes.

\section{Science results}
\label{science}
We constructed an astronomical catalog for the entire 55 TB SDSS dataset.
This catalog is the first fully
Bayesian astronomical catalog for a modern astronomical imaging survey.

Absolute truth is unknowable for astronomical catalogs, but validation
is nonetheless essential.
One area of the sky (Stripe 82) has been imaged approximately 80
times in SDSS, whereas most regions have been imaged just once.
This region provides a
convenient validation strategy: combine exposures
from all Stripe-82 runs to produce a very high signal-to-noise image, and estimate ground
truth parameters from that image.

``Photo''~\cite{lupton2005sdss} is a state-of-the-art software pipeline for
constructing large astronomical catalogs.  Photo is a carefully hand-tuned
heuristic. We use Photo's estimated parameters from the combined Stripe 82
imagery as ground truth. We then compare Photo's output for a single run's
imagery to Celeste's output for the same images. Although this ``ground truth'' is
still prone to errors, such errors typically favor Photo, since any systematic
errors will be consistent in Photo's output.

In addition to point estimates, Celeste offers measures of posterior uncertainty
for source type (star or galaxy), brightness, and colors. This is a novel
feature owing to Celeste's formulation as a Bayesian model, offering
astronomers a principled measure of the quality of inference for each light
source. No such analogue exists for Photo.

\begin{table}
\caption{Average error based on light sources from Stripe 82.}

\centering
\renewcommand{\arraystretch}{1.1}
\begin{tabular}{lrr}

\hline
\rule{0pt}{8pt}        & Photo & Celeste \\

\hline
\rule{0pt}{5pt}Position     & 0.36     & \textbf{0.27}\\
Missed gals  & \textbf{0.06} & 0.19\\
Missed stars & 0.12  & 0.15\\
Brightness   & 0.21     & \textbf{0.14} \\
Color u-g    & 1.32     & \textbf{0.60} \\
Color g-r    & 0.48     & \textbf{0.21} \\
Color r-i    & 0.25     & \textbf{0.12} \\
Color i-z    & 0.48     & \textbf{0.17} \\
Profile      & 0.38     & \textbf{0.28} \\
Eccentricity & 0.31     & \textbf{0.23} \\
Scale        & 1.62     & 0.92 \\
Angle        & 22.54     & \textbf{17.54} \\

\hline
\end{tabular}

\vspace{10px}
\begin{flushleft}
  	\textbf{Lower is better.} Results in bold are better by more than two standard deviations.
    ``Position'' is error, in pixels, for the location of the light sources'
    centers.
    ``Missed gals'' is the proportion of galaxies labeled as stars.
    ``Missed stars'' is the proportion of stars labeled as galaxies.
    ``Brightness'' measures the reference band (r-band) magnitude.
    ``Colors'' are ratios of magnitudes in consecutive bands.
    ``Profile'' is a proportion indicating whether a galaxy is de Vaucouleurs
    or exponential.
    ``Eccentricity'' is the ratio between the lengths of a galaxy's minor and
    major axes.
    ``Scale'' is the effective radius of a galaxy in pixels.
    ``Angle'' is the orientation of a galaxy in degrees.
\end{flushleft}
\vspace{-1em}

\label{scientific_results}
\end{table}

The first numeric column of Table~\ref{scientific_results} shows error on each
parameter from Photo, averaged over all sources in the chosen image. The second
numeric column shows results for Celeste. Celeste achieves improved accuracy on
nearly all estimated parameters, usually by a substantial margin, improvements
that are both statistically significant and of practical significance to
astronomers. Photo outperforms Celeste only at classifying galaxies. Upon manual
inspection, the true class of these mislabeled light sources often seem to be
genuinely ambiguous. Some may be quasars, which are collapsing galaxies with
point-like spatial characteristics of stars. Celeste's posterior uncertainty
reflects the ambiguity.

\section{Discussion}
\label{discussion}

\subsection{Bayesian inference and distributed optimization}

Our work demonstrates conclusively that massive, complex datasets can be analyzed with Bayesian inference. It is not necessary to resort to a heuristic when Bayesian inference is preferable, or to settle for recovering just the mode of the posterior. Through a combination of algorithmic techniques that were previously unexplored for variational inference, and a careful implementation, high throughput is possible.

Distributed optimization is important for variational inference, as well as for optimization problems that do not arise from statistical models. The multi-level approach we propose is well-suited to HPC numerical optimization because it utilizes all available levels of parallelism. It has only modest communication requirements, yet converges in few iterations to a stationary point.

\subsection{Convergence of high-productivity languages and high-performance languages}

Celeste constitutes the first successful use of Julia at petascale, demonstrating Julia's viability for HPC applications where performance constraints would have otherwise demanded the use of a low-level language like C or Fortran. Our experience challenges the traditional HPC wisdom that
interactive, high-level languages are unsuitable for performance hot spots, and
instead must be limited to a ``glue'' role, such that the majority of an
application's runtime is spent in low-level kernels.

Julia may not always be preferable to a traditional HPC workflow---legacy code, existing libraries, and the time required to learn a new language are all good reasons why C and Fortran will remain popular for HPC.
But the reasons why Julia worked well for Celeste seem far from
unique to our project.
New HPC projects---particularly those planning to write prototype code and production code in different languages---may benefit from using Julia instead.

\section*{Acknowledgments}

This research used resources of the National Energy Research Scientific Computing Center, a DOE Office of Science User Facility supported by the Office of Science of the U.S. Department of Energy under Contract No.~DE-AC02-05CH11231.

The authors are grateful to the NERSC staff, especially the computational systems group:
David Paul, James Botts, Scott Burrow, Tina Butler, Tina Declerck, Douglas Jacobsen, Jay Srinivasan, and Bhupender Thakur.
Without their support these results would not have been possible.
We also thank Brandon Cook and Rebecca Hartman-Baker for coordinating our runs on Cori Phase~II.
We thank Brian Friesen for providing real-time measurements of node health.
Brian Austin, Deborah Bard, Wahid Bhimji, Lisa Gerhardt, Glenn Lockwood, Quincey Koziol, and Jialin Liu provided valuable advice on effective use of Lustre and Burst Buffer file systems.

We owe an enormous debt to the Julia and LLVM open source communities. In particular, we thank Jeff Bezanson, Jameson Nash and Yichao Yu.

The authors are grateful for funding from the MIT Deshpande Center for Numerical Innovation, the Intel Technology Science Center for Big Data at MIT, DARPA Xdata, the Singapore MIT Alliance, NSF Award DMS-1016125, a DOE grant with Dr. Andrew Gelman of Columbia University for petascale hierarchical modeling, Aramco Oil due to Ali Dogru, and the Gordon and Betty Moore Foundation.

We thank Alan Edelman, Viral Shah, Bharat Kaul, and Pradeep Dubey for supporting this project.

\bibliographystyle{IEEEtran}
\bibliography{IEEEabrv,references}

\begin{thebibliography}{10}
\providecommand{\url}[1]{#1}
\csname url@samestyle\endcsname
\providecommand{\newblock}{\relax}
\providecommand{\bibinfo}[2]{#2}
\providecommand{\BIBentrySTDinterwordspacing}{\spaceskip=0pt\relax}
\providecommand{\BIBentryALTinterwordstretchfactor}{4}
\providecommand{\BIBentryALTinterwordspacing}{\spaceskip=\fontdimen2\font plus
\BIBentryALTinterwordstretchfactor\fontdimen3\font minus
  \fontdimen4\font\relax}
\providecommand{\BIBforeignlanguage}[2]{{%
\expandafter\ifx\csname l@#1\endcsname\relax
\typeout{** WARNING: IEEEtran.bst: No hyphenation pattern has been}%
\typeout{** loaded for the language `#1'. Using the pattern for}%
\typeout{** the default language instead.}%
\else
\language=\csname l@#1\endcsname
\fi
#2}}
\providecommand{\BIBdecl}{\relax}
\BIBdecl

\bibitem{dr12}
S.~Alam, F.~D. Albareti \emph{et~al.}, ``The eleventh and twelfth data releases
  of the sloan digital sky survey: Final data from {SDSS-III},'' \emph{The
  Astrophysical Journal Supplement Series}, vol. 219, no.~1, p.~12, 2015.

\bibitem{LSST}
{Large Synoptic Survey Telescope Consortium}, ``About {LSST},''
  \url{http://www.lsst.org/about}, 2017, [Online; accessed September 12, 2017].

\bibitem{bezanson2017julia}
J.~Bezanson, A.~Edelman, S.~Karpinski, and V.~B. Shah, ``Julia: A fresh
  approach to numerical computing,'' \emph{SIAM Review}, vol.~59, no.~1, pp.
  65--98, 2017.

\bibitem{lupton2005sdss}
R.~H. Lupton, Z.~Ivezic \emph{et~al.}, ``{SDSS} image processing {II}: The
  photo pipelines,'' Technical report, Princeton University, Tech. Rep., 2005.

\bibitem{bertin1996sextractor}
E.~Bertin and S.~Arnouts, ``{SExtractor}: Software for source extraction,''
  \emph{Astronomy and Astrophysics Supplement Series}, vol. 117, no.~2, pp.
  393--404, 1996.

\bibitem{bishop2006pattern}
C.~M. Bishop, \emph{Pattern Recognition and Machine Learning}.\hskip 1em plus
  0.5em minus 0.4em\relax Springer, 2006.

\bibitem{thetractor}
D.~Lang, ``{The Tractor},'' \url{http://thetractor.org/about/}, 2017, [Online;
  accessed October 22, 2017].

\bibitem{alexanderian2016fast}
A.~Alexanderian, N.~Petra, G.~Stadler, and O.~Ghattas, ``A fast and scalable
  method for {A}-optimal design of experiments for infinite-dimensional
  {B}ayesian nonlinear inverse problems,'' \emph{SIAM Journal on Scientific
  Computing}, vol.~38, no.~1, pp. A243--A272, 2016.

\bibitem{blei2017variational}
D.~M. Blei, A.~Kucukelbir, and J.~D. McAuliffe, ``Variational inference: A
  review for statisticians,'' \emph{Journal of the American Statistical
  Association}, 2017.

\bibitem{hoffman2013stochastic}
M.~D. Hoffman, D.~M. Blei, C.~Wang, and J.~W. Paisley, ``Stochastic variational
  inference.'' \emph{Journal of Machine Learning Research}, vol.~14, no.~1, pp.
  1303--1347, 2013.

\bibitem{ousterhout1998scripting}
J.~K. Ousterhout, ``Scripting: Higher level programming for the 21st century,''
  \emph{Computer}, vol.~31, no.~3, pp. 23--30, 1998.

\bibitem{walt2011numpy}
S.~v.~d. Walt, S.~C. Colbert, and G.~Varoquaux, ``The {NumPy} array: A
  structure for efficient numerical computation,'' \emph{Computing in Science
  \& Engineering}, vol.~13, no.~2, pp. 22--30, 2011.

\bibitem{vincent2016pyfr}
P.~Vincent, F.~Witherden, B.~Vermeire, J.~S. Park, and A.~Iyer, ``Towards green
  aviation with python at petascale,'' in \emph{Proceedings of the
  International Conference for High Performance Computing, Networking, Storage
  and Analysis}, ser. SC'16, 2016.

\bibitem{lattner2004llvm}
C.~Lattner and V.~Adve, ``{LLVM}: A compilation framework for lifelong program
  analysis \& transformation,'' in \emph{Proceedings of the International
  Symposium on Code Generation and Optimization}.\hskip 1em plus 0.5em minus
  0.4em\relax IEEE Computer Society, 2004, p.~75.

\bibitem{regier2015celeste}
J.~Regier, A.~Miller, J.~McAuliffe, R.~Adams, M.~Hoffman, D.~Lang, D.~Schlegel,
  and Prabhat, ``Celeste: Variational inference for a generative model of
  astronomical images,'' in \emph{Proceedings of the 32nd International
  Conference on Machine Learning}, 2015.

\bibitem{recht2011hogwild}
B.~Recht, C.~Re, S.~Wright, and F.~Niu, ``Hogwild: A lock-free approach to
  parallelizing stochastic gradient descent,'' in \emph{Advances in Neural
  Information Processing Systems}, 2011.

\bibitem{pamnany2015dtree}
K.~Pamnany, S.~Misra, V.~Md, X.~Liu, E.~Chow, and S.~Aluru, ``Dtree: Dynamic
  task scheduling at petascale,'' in \emph{International Conference on High
  Performance Computing}.\hskip 1em plus 0.5em minus 0.4em\relax Springer,
  2015, pp. 122--138.

\bibitem{pgas}
{Tim Stitt}, ``An introduction to the partitioned global address space
  programming model.''

\bibitem{nieplocha2006advances}
J.~Nieplocha, B.~Palmer, V.~Tipparaju, M.~Krishnan, H.~Trease, and E.~Apr{\`a},
  ``Advances, applications and performance of the {Global Arrays Shared Memory
  Programming Toolkit},'' \emph{International Journal of High Performance
  Computing Applications}, vol.~20, no.~2, pp. 203--231, 2006.

\bibitem{hoefler2015rma}
T.~Hoefler, J.~Dinan, R.~Thakur, B.~Barrett, P.~Balaji, W.~Gropp, and
  K.~Underwood, ``Remote memory access programming in {MPI-3},'' \emph{ACM
  Trans. Parallel Comput.}, vol.~2, no.~2, pp. 9:1--9:26, Jun. 2015.

\bibitem{wright1999numerical}
S.~Wright and J.~Nocedal, ``Numerical optimization,'' \emph{Springer Science},
  vol.~35, pp. 67--68, 1999.

\bibitem{pan2016cyclades}
X.~Pan, M.~Lam, S.~Tu, D.~Papailiopoulos, C.~Zhang, M.~I. Jordan,
  K.~Ramchandran, and C.~R\'{e}, ``Cyclades: Conflict-free asynchronous machine
  learning,'' in \emph{Advances in Neural Information Processing Systems},
  2016.

\bibitem{staticarrays}
{Andy Ferris}, ``{StaticArrays.jl},''
  \url{https://github.com/JuliaArrays/StaticArrays.jl}, 2017, [Online; accessed
  October 22, 2017].

\bibitem{structsofarrays}
{Simon Kornblith}, ``{StructsOfArrays.jl},''
  \url{https://github.com/simonster/StructsOfArrays.jl}, 2017, [Online;
  accessed September 12, 2017].

\bibitem{revels2016forwarddiff}
J.~Revels, M.~Lubin, and T.~Papamarkou, ``Forward-mode automatic
  differentiation in {J}ulia,'' in \emph{Proceedings of the 7th International
  Conference on Algorithmic Differentiation}, 2016.

\bibitem{reversediff}
{Jarrett Revels}, ``{ReverseDiff.jl},''
  \url{https://github.com/JuliaDiff/ReverseDiff.jl}, 2017, [Online; accessed
  September 12, 2017].

\bibitem{feng2016sparsehessian}
F.~Qiang, C.~Petra, M.~Lubin, J.~Huchette, and M.~Anitescu, ``On efficient
  {H}essian computation using the edge pushing algorithm in {J}ulia,'' in
  \emph{Proceedings of the 7th International Conference on Algorithmic
  Differentiation}, 2016.

\bibitem{Bhimji2016AcceleratingSW}
W.~Bhimji, D.~Bard, M.~Romanus, D.~Paul \emph{et~al.}, ``Accelerating science
  with the {NERSC Burst Buffer Early User Program},'' \emph{Cray User Group
  2016 Proceedings}, 2016.

\end{thebibliography}

\end{document}